\documentclass{aastex61}

\newcommand\aastex{AAS\TeX}

\received{July 1, 2016}
\revised{September 27, 2016}
\accepted{\today}
\submitjournal{ApJ}

\shorttitle{\aastex\ sample article}
\shortauthors{Centeno et al.}

\begin{document}

\title{On the (mis)Interpretation of the scattering polarization
  signatures in the Ca {\sc II} 8542 \AA\ line through spectral line inversions}

\correspondingauthor{Rebecca Centeno}
\email{rce@ucar.edu}

\author[0000-0002-1327-1278]{Rebecca Centeno}
\affil{High Altitude Observatory (NCAR), 3080 Center Green Dr., Boulder, CO, USA}
\author[0000-0002-4640-5658]{Jaime de la Cruz Rodr\'\i guez}
\affil{Institute for Solar Physics, Dept. of Astronomy, Stockholm
  University, AlbaNova University Centre, SE-10691 Stockholm, Sweden.}
\author{Tanaus\'u del Pino Alem\'an}
\affil{Instituto de Astrof\'\i sica de Canarias, 
  C. V\'\i a L\'actea s/n, La Laguna, Tenerife, Spain.}

\begin{abstract}
Scattering polarization tends to dominate the linear polarization
signals of the Ca II 8542 \AA\ line in weakly magnetized areas ($B
\lesssim 100$ G), especially when
the observing geometry is close to the limb. In
this paper we evaluate the degree of applicability of existing non-LTE
spectral line inversion codes (which assume that the spectral line
polarization is due to the Zeeman effect only) at inferring the
magnetic field vector and, particularly, its transverse
component. To this end, we use the inversion code STiC
to extract the strength and orientation of the
magnetic field from synthetic spectropolarimetric data generated with the Hanle-RT code. The latter accounts for the generation of polarization through scattering processes as well as the joint actions of the Hanle and the Zeeman effects. 
We find that, when the transverse component of the field is stronger
than $\sim$80 G, the inversion code is able to retrieve accurate
estimates of the transverse field strength as well as its azimuth in the plane of the sky. Below this threshold, the scattering polarization signatures become the major contributors to the linear polarization signals and often mislead the inversion code into severely over- or under-estimating the field strength.
Since the line-of-sight component of the field is derived from the circular polarization signal, which is not affected by atomic alignment, the corresponding inferences are always good.

\end{abstract}
\keywords{solar magnetic fields}

\section{Introduction} \label{sec:intro}

Spectropolarimetry, the science of measuring the intensity and
polarization of light as a function of wavelength, is a very
powerful tool for remote sensing of thermodynamic and
magnetic properties of astrophysical plasmas.

There are several physical mechanisms that can generate polarization
in spectral lines. They are all related to the lack of symmetry in the
absorption and excitation mechanisms of the atom. 
The mere presence of spectral line polarization tells us
something about the conditions of the plasma where the light was
emitted or absorbed. The subtleties of the intensity and polarization spectra encode a wealth of information about how the atoms were excited, and thus about the conditions dominating the atmosphere where the observed radiation was generated.

A spectral line arises from the absorption or emission of a photon
when an electron transitions from one bound atomic (or molecular) energy level to another. These
energy levels are, in general, degenerate. In the presence of an
external magnetic field, these magnetic sub-levels (represented by the
magnetic quantum number $M$) split and the
degeneracy is broken, leading to a wavelength shift between the $\pi$
($\Delta M = 0$) and $\sigma$ ($\Delta M = \pm 1$) components of the
spectral line. The splitting of the energy levels is, in general, proportional to
the magnetic field strength, and this polarization mechanism is commonly known as the Zeeman effect. 
If the magnetic field is very weak (relative to the width of the
spectral line), or absent, the wavelength shift of the
magnetic sub-levels is very small and
the ensuing polarization signatures are negligible. This is especially
true for the linear polarization signatures of chromospheric lines, which scale with the square of the ratio between the Zeeman splitting
and the line widths, instead of linearly, as the circular polariztion
features do.

\noindent But even in the absence of magnetic fields, there are other mechanisms
that can generate polarization in spectral lines. When the magnetic
sub-levels of a given energy level are unevenly populated (i.e. atomic level
polarization), this results in an
imbalance between the number of $\pi$ and $\sigma$ transitions per
unit volume and time, which leads to emergent polarization
signatures without the need to invoke a wavelength shift (see
\cite{trujillobueno_review} for a
short introductory review to scattering polarization and the Hanle
effect and \cite{landi_book} for a detailed explanation of the origin of
polarization in spectral lines). The population imbalance and quantum
coherence among the degenerate magnetic sublevels, known as atomic
level polarization, lead to an imbalance between
the rates of different $|\Delta M|$ transitions, which, in turn, produces
emergent linear polarization signals, commonly referred to as
``scattering polarization''.

In the Sun's atmosphere, the most common mechanism for creating 
atomic level polarization is the so-called optical pumping, that is, the
anisotropic illumination of the atoms leads to selective emission or absorption of the
different spectral line components, and to scattering linear polarization, without the need for a magnetic field.
Elastic collisions have a strong depolarizing effect, and tend to
equalize the populations of the magnetic sublevels. So for this
mechanism to be effective, the radiative transitions have to dominate
the atomic excitation and deexcitation processes. In the rarefied environment of the
chromosphere, collisions are infrequent and the radiation is weakly
coupled to the local thermodynamic conditions (this is commonly
referred to as non Local Thermodynamic Equilibrium or non-LTE). This sets a perfect stage for
optical pumping to act effectively, creating atomic level polarization
that leads to the generation of polarization signatures in spectral
lines in the absence of an external magnetic or electric field.
In chromospheric quiet Sun regions, where the magnetic fields are weak and the
Zeeman effect generates small polarization signatures, the
linear polarization of chromospheric lines is typically dominated by scattering
processes  \cite[][]{mansosainz2010, mansosainz2003}.

The Hanle effect is the relaxation of the atomic level coherences due to
the presence of a magnetic field. When a magnetic
field is present, the emergent scattering polarization signals will change with
respect to the zero-field case, leaving a measurable imprint on the
polarization spectrum.
The Hanle effect is typically sensitive to relatively weak magnetic field strengths (depending on the
spectral line) and allows us to access regimes where the Zeeman effect does not
have useful diagnostic capabilities.

The Ca {\sc ii} 8542 \AA\ line has been deemed one of the most
promising lines for chromospheric magnetic field diagnostics \cite[][]{lagg_review,
  quinteronoda, quinteronoda2} due to its
accessibility from ground-based observations and the relative ease of
its interpretation -- this is, when modeling it, it is safe to assume complete redistribution
\cite[][]{uitenbroek1989}, ignore non-equilibrium ionization effects
\cite[][]{wedemeyer}, and full 3-D radiative transfer computations are unnecessary in
certain scenarios \cite[][]{jaime2012}.
{\bf It is of interest to note that neglecting 3D radiative transfer
  effects can have a large impact on the scattering polarization
  signals of this and other spectral lines \citep[see][for an analysis
  of their impacts on Ca {\sc i} 4227\AA]{ca4227}. The presence of horizontal inhomogeneities in the physical
  properties of the Sun's atmosphere can also break the axial symmetry
  of the radiation field, fueling an additional source of atomic level
  polarization. \citet{stepan}
  quantified the error in the emergent linear polarization of Ca {\sc
    ii} 8542 \AA\ when
  neglecting said horizontal inhomogeneities, and concluded that it
  could be as large as the linear polarization amplitudes themselves. This
  study was carried out on the polarization signals emerging from a
  high resolution quiet Sun radiation magneto-hydrodynamic simulation,
  and probably represents an upper bound for these scattering
  polarization signals. Spatial degradation of the spectral profiles and
  stronger magnetic fields will likely render
these 3-D radiative transfer effects less important.}

Despite it being commonly used to extract information about the
chromospheric magnetic field, whether the interpretation is
done by applying the weak field approximation \cite[WFA, see, for
instance][]{centenoWFA, kuridze, morosin2020} or using non-LTE spectral line inversion codes
\cite[][]{stic, nicole, snapi, desire}, these diagnostic tools do not account for the physics of scattering polarization nor
its modification via the Hanle effect. However, scattering polarization
signatures typically dominate
the linear polarization profiles of Ca {\sc ii} 8542 \AA\ in weak
field areas \cite[][]{mansosainz2010, mansosainz2003}, particularly close to the
limb, where the observing geometry maximizes these polarization signatures.
When combined with the enhancing effect of shocks, the
linear polarization signatures can reach amplitudes of up to 1\% of the
continuum intensity \cite[][]{carlin2012, carlin2013}, which would dominate over
Zeeman signatures induced by magnetic fields in the hecto-gauss range.

To date, the only spectral line inversion codes that account for the
physics of scattering polarization and the combined action of the
Hanle and Zeeman effects are limited to the interpretation of some
neutral He lines, where computing the
radiative transfer in a slab of constant properties can be a
suitable approximation \cite[see, for instance][]{hazel}.

This work attempts to evaluate the inaccuracy in the chromospheric
magnetic field values retrieved from non-LTE inversions of the Ca {\sc
  ii} 8542 \AA\ line when neglecting the scattering
polarization signals and their modification due to the Hanle effect.
Section \ref{sec:methodology} describes
the computer codes and numerical setups used to synthesize and invert
the spectral lines. In section \ref{sec:analysis} we present the
analysis of the inversion results, followed by some brief thoughts and conclusions in
section \ref{sec:conclusions}.

\section{Synthesis and inversion of spectral lines}\label{sec:methodology}

Purely a numerical experiment, this work aims at quantifying the
biases incurred by non-LTE spectral line inversion codes when 
analyzing Ca {\sc ii} 8542 \AA\ Stokes spectra.
Currently, the non-LTE spectral line inversion codes capable of
interpreting the spectral line radiation in the Ca {\sc ii} IR triplet
do not account for the generation of polarization induced by scattering
processes and its modification due to the Hanle effect.
There are, however, forward modeling radiative transfer software packages that
account for the subtle quantum-mechanical effects responsible for
these processes, which can realistically model the polarization signatures of Ca
{\sc ii} 8542 \AA.

\noindent Hanle-RT \cite[see, for instance,][]{hanlert} is a 1.5-D
radiative transfer code able to solve the polarized radiation transfer
problem in a plane-parallel geometry, under non-LTE conditions, for a multi-term or multi-level atom in the presence of an arbitrary magnetic field, taking
into account the effects of partial frequency redistribution (PRD), as well
as the contribution of inelastic and elastic collisions.
{\bf The rate of depolarizing elastic collisions with neutral hydrogen are the result of ab initio quantum mechanical calculations by \citep{elastic_collisions}}. Spectral line
polarization is generated by means of scattering processes as
well as by the combined action of the Hanle and Zeeman effects.

\subsection{Synthesis}

\begin{figure}[!t]
\includegraphics[angle=0,scale=.7]{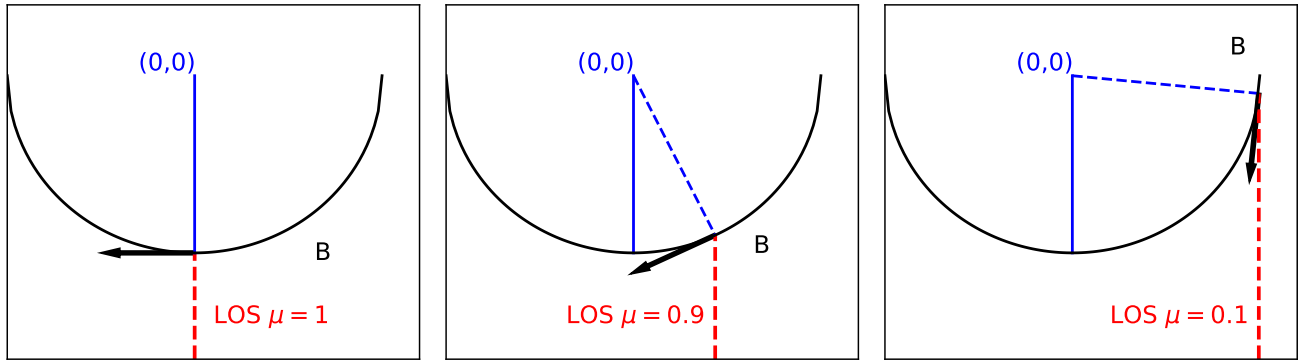}
\caption{Representation of three of the five LOS geometries used in the
  Hanle-RT forward modeling calculations ($\mu= 1$, $\mu=0.9$ and $\mu
  = 0.1$ shown here). This figure shows a top view of a cross-section through
  the Sun's equator, where (0,0) represents the center of the Sun. The
  magnetic field vector, $\vec B$,  is always tangent to the solar
  surface, and its LOS component points towards the observer for the observing
  geometries away fom the disk center. The Sun's
  equator, the magnetic field and the LOS are in the same plane.  \label{fig:geometry}}
\end{figure}

The Hanle-RT code was used to synthesize the Stokes profiles of the Ca
{\sc ii} 8542 \AA\ line in a number of magnetic field scenarios and
observing geometries. The spectra were synthesized in the 1-D 
semi-empirical C model atmosphere of \cite{falc}, hereafter, FALC. Constant magnetic
fields with varying strengths (from 0 to 100 G in steps of 20 G and
from 100 G to 1000 G in steps of 100 G), horizontal with respect to the local solar surface,
were introduced ad-hoc in the model to generate the corresponding
Zeeman/Hanle polarization signatures.
The spectra were generated for lines-of-sight (LOS) with five different
heliocentric angles ($\theta$), from
disk center to close to the limb ($\mu = 1, 0.9, 0.6, 0.5, 0.1$, where
$\mu = {\rm cos} \theta$). In all geometries, the magnetic field was kept horizontal to the local solar surface, so the angle between the magnetic
field direction and the LOS adopts different values (and so does the
ratio of linear to circular Zeeman-induced polarization from the
observer's perspective). Figure \ref{fig:geometry} depicts three of the LOS
geometries listed above, where the black arrow represents the magnetic
field vector, always tangential to the solar surface, and the red
dashed line shows the line-of-sight.

The atomic model of Ca {\sc ii} used in this work has
five bound energy levels as well as the Ca {\sc iii} continuum, allowing for
5 bound-bound radiative transitions \cite[i.e. the Ca {\sc ii} H \& K lines in
the near UV and the Ca {\sc ii} infrared triplet, see Fig. 1
in][]{centenoWFA}. 
Ca {\sc ii} 8542 \AA\ is a magnetically sensitive line that arises from the
transition between the $3^2D_{5/2}$ and the $4^2P_{3/2}$ levels.
Both in the Hanle-RT synthesis as well as in the spectral line inversion, the IR
triplet is computed assuming complete frequency redistribution (CRD),
whilst the H \& K UV lines are treated in PRD accounting for Raman scattering (also referred to
as ``cross-redistribution'' in some works) from the IR triplet to the
UV doublet. 
The spectral lines were synthesized on a regular wavelength grid with a
spectral sampling of 1 m\AA\ and a coverage that spanned $\pm 1.2$ \AA\
from the line core. No additional broadening mechanisms
(aside from the natural, thermal and collisional broadenings) were included in the calculations, which
allowed for a fair amount of continuum in the polarization spectra.

Two sets of synthetic Stokes spectra were computed for each magnetic
field strength and each observing geometry, one
in which the linear polarization signatures were a consequence of
scattering phenomena as well as the joint actions of the Hanle and
Zeeman effects, and another one in which the atomic level polarization was
``turned off'' and only the signals due to the Zeeman effect were permitted.
This allowed us to evaluate the performance of the spectral line
inversion of purely Zeeman-induced signatures against the inversion of
the more realistic scenario including scattering polarization and the
Hanle effect. The total number of synthetic Stokes profiles is 150 (15
values of the field strength for 5 observing geometries and 2 calculation
modes - with and without atomic level polarization).

\begin{figure}
\includegraphics[angle=0,scale=.49]{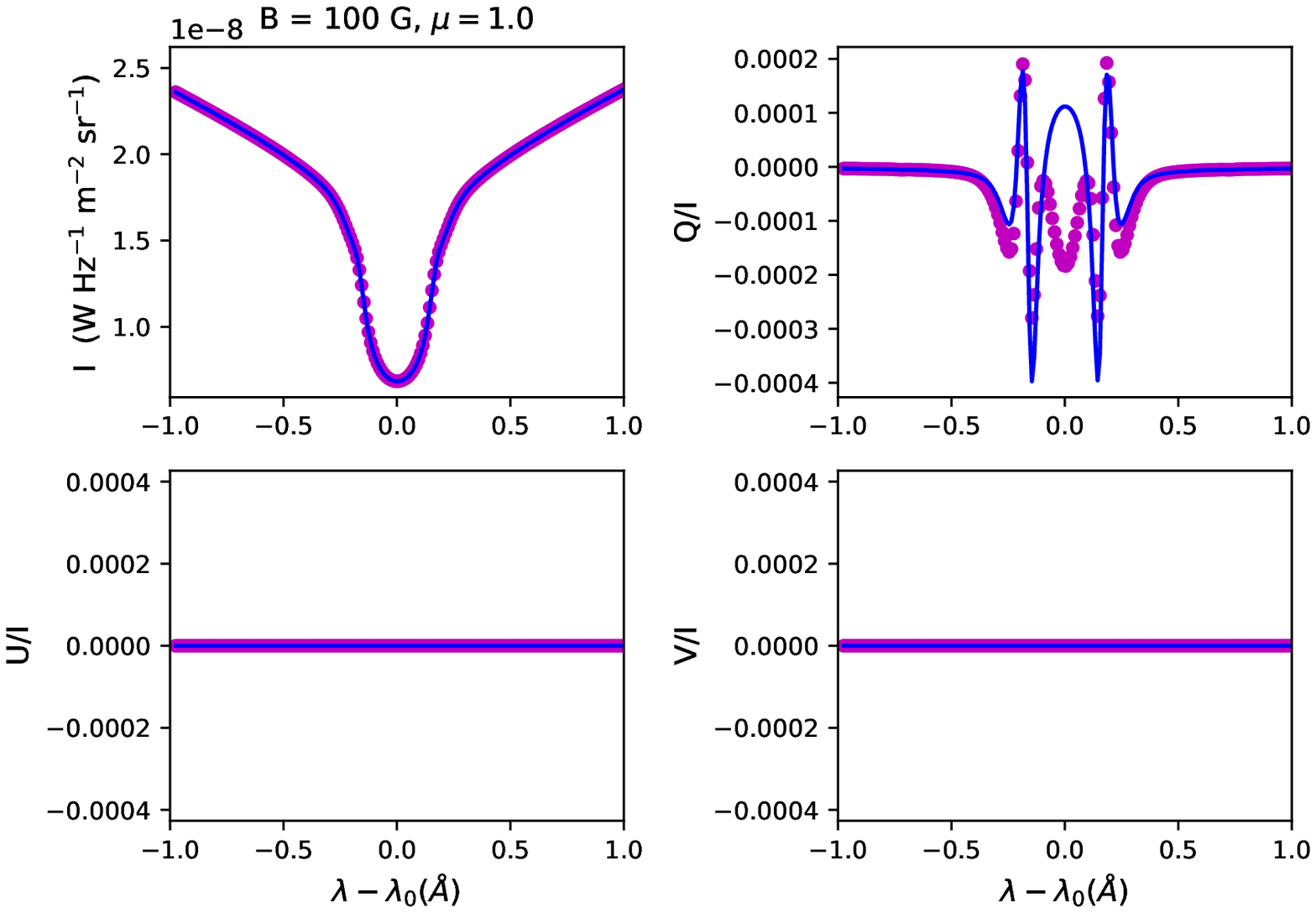}
\includegraphics[angle=0,scale=.49]{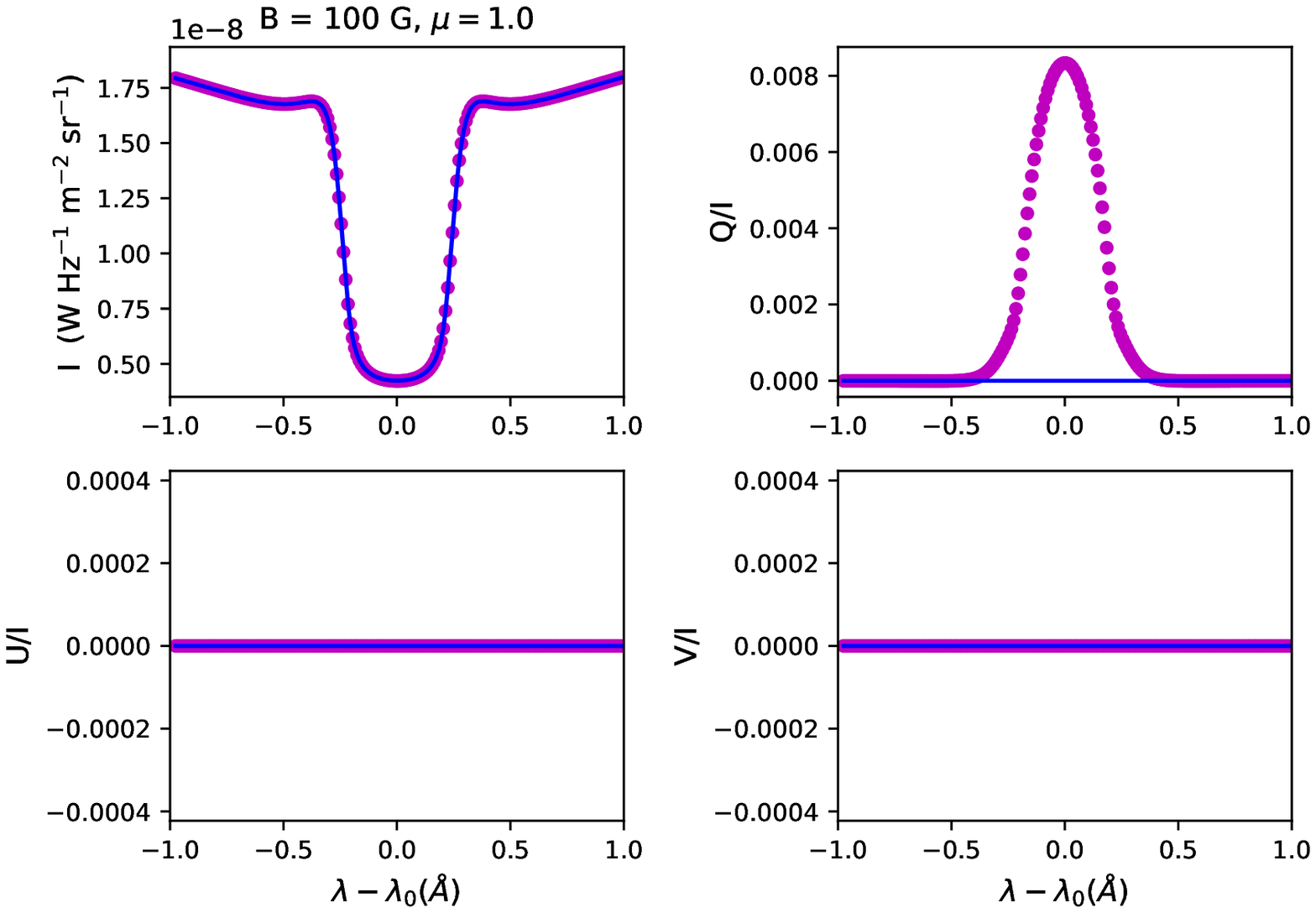}\\
\includegraphics[angle=0,scale=.49]{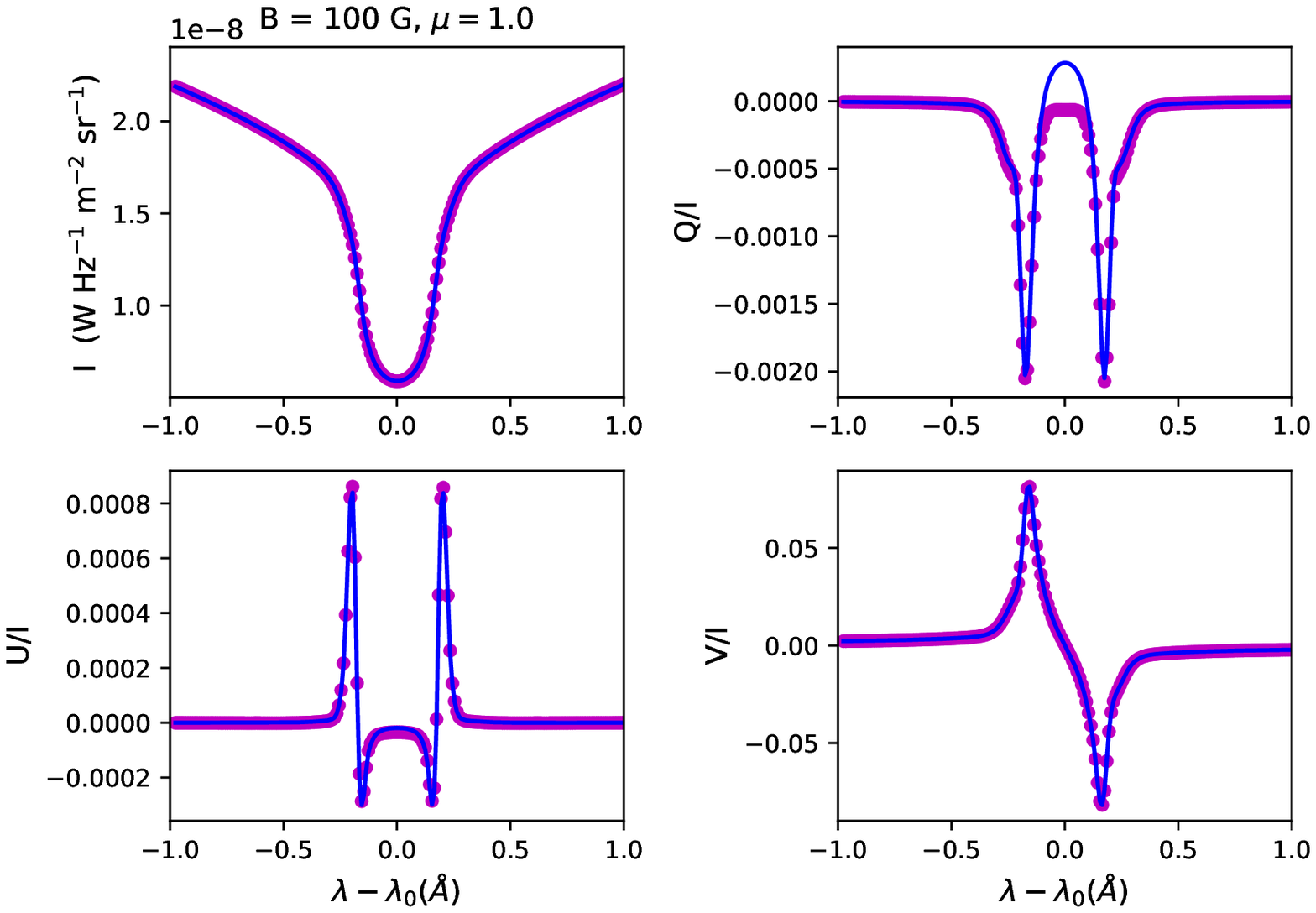}
\includegraphics[angle=0,scale=.49]{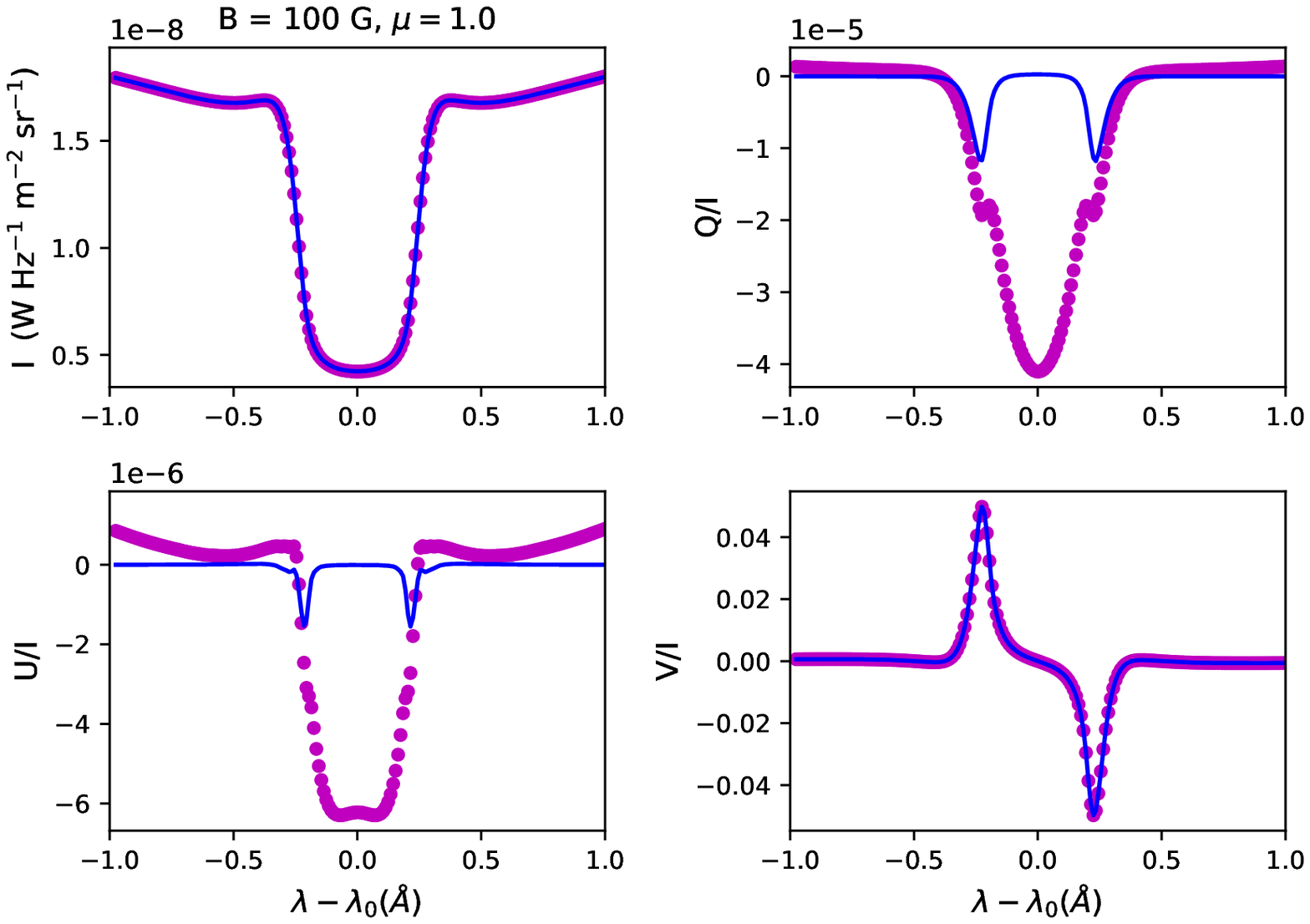}
\caption{Examples of sets of synthetic Stokes profiles for different magnetic
  field strengths and different observing geometries. The magenta dots
  correspond to the profiles computed taking into account scattering
  polarization and the joint action of the Hanle and Zeeman effects,
  whilst the blue solid line corresponds to the calculations where
  only Zeeman-induced polarization was permitted. The field strength
  and the observing geometry, $\mu$, are labeled at the top of each Stokes I panel.
\label{fig:profiles}}
\end{figure}

Figure \ref{fig:profiles} shows a few examples of the synthetic Stokes
profiles for several field strengths and observing geometries. As
expected, the Stokes spectra for the LOS with $\mu=0.1$ (the two examples on the
right side of the figure) show the largest differences
between the Zeeman-only case (blue solid line) and the full calculation with scattering
polarization (magenta dots). This is because 
observing geometries very close to the limb favor larger
contributions of scattering polarization to the Stokes profiles.
Note that for the disk center geometry ($\mu=1.0$), V/I is always
zero, because the magnetic field is tangential to the surface and
therefore transverse to the line-of-sight (see Fig. \ref{fig:geometry}). In all other geometries,
there is a V/I signal as long as the magnetic field is
non-zero. Stokes I and Stokes V are not affected by this form of
atomic level polarization (atomic alignment), so the Zeeman-only and
full calculation profiles look identical.

\subsection{Inversion}

As stated earlier, there is currently no non-LTE spectral line
inversion code capable of treating the physical
processes that lead to atomic level polarization and the Hanle effect
in the Ca {\sc ii} 8542 \AA\ line.
In order to assess how ``standard'' non-LTE spectral line inversion codes interpret
the polarized signatures due to these subtle quantum-mechanical effects, we
inverted the synthetic Hanle-RT profiles with the STockholm
inversion Code \cite[STiC,][]{stic, stic_orig} and compared the
inversion results to the magnetic field of the model used to generate
the original spectra.

STiC is a parallelized non-LTE full-Stokes spectral line inversion 
code that uses a modified version of
the Rybycki and Hummer radiative transfer code \cite[RH,][]{rh} to
solve for the atomic population densities assuming statistical
equilibrium in a plane-parallel atmosphere. 
The code can account for PRD effects
\cite[][]{PRD}, which are important for the modeling of many chromospheric lines. 
The polarized radiative transfer equation is solved using cubic Bezier solvers
\cite[][]{DELO}. The inversion engine of STiC is based on a
Levenberg-Marquardt algorithm \cite[][]{numerical_recipes}, and includes an equation of state extracted
from the Spectroscopy Made Easy \cite[SME,][]{piskunov2017} code. Polarization is exclusively generated by the Zeeman effect,
therefore STiC cannot account for polarization due to
scattering processes and its modification via the Hanle effect.

As in the synthesis, the Ca {\sc ii} model atom used in the inversions
contains 5 bound energy levels and the Ca {\sc iii} continuum, which
allows for the bound-bound radiative transitions of Ca {\sc ii} H \& K and the
Ca IR triplet. As in the previous section, the H \& K lines are
treated in the PRD regime, while the IR triplet is computed assuming
CRD.

The inversions were carried out in two cycles, following the strategy
of \cite{[][]vissers2021}, with 4 nodes for the temperature in the first
cycle and 7 in the second cycle. The number of nodes for the rest of
the parameters stayed constant throughout the 2 cycles, with 3 for the
microturbulent velocity, and one for each one of the remaining
physical parameters (the LOS velocity, the LOS component of
the magnetic field, the transverse component of the magnetic field and
its azimuth in the plane of the sky). In the case of the viewing angle
with $\mu=0.1$, additional nodes for the temperature and the
microturbulent velocity were needed in the second cycle in order to obtain a good fit to
the Stokes I profile. However, STiC implements a regularization of the merit function
that minimizes artifacts in the atmospheric gradients that result from the
excess freedom in the parameters of the model atmosphere
\cite[see][for details]{stic}.

The initial guess model atmosphere was the same for all inversions. It
consisted of a version of the FAL-C model with a perturbed temperature
profile and a constant value for the LOS velocity and the magnetic field vector.
Each set of Stokes spectra was paired with a set of weights to ensure
a good inversion fit for all Stokes parameters simultaneously \cite[][]{inversion_review}. The inversion
weights are a set of 4 values used to emphasize or de-emphasize the
relative importance of a given Stokes parameter with respect to Stokes
I in the evaluation of the merit function, $\chi^2$, that the inversion algorithm
is trying to minimize \cite[see explanation in][for instance]{centenoHMI}:

\begin{equation}
\chi^2 =  \sum_s \frac{w_s^2}{\sigma_s}\sum_{\lambda} [I^{\rm OBS}_s(\lambda) -
 I^{\rm SYN}_s(\lambda)]^2
\end{equation}

\noindent where the weights, $w_s$, were chosen to
approximately equalize the amplitudes of the synthetic Hanle-RT Stokes Q, U and V to that of
Stokes I. In this expression, $I^{\rm OBS}$ represents the Hanle-RT Stokes spectra to be
inverted and $I^{\rm SYN}$ represents the best fit found by the STiC
inversion code. $\sigma_s$ is the expected noise of each one of the Stokes
parameters. The data that we were inverting were noiseless, so we
artificially set to $\sigma_s \sim 10^{-3}$ times the continuum
intensity to indicate to STiC the magnitude of the acceptable error in fitting 
Stokes I. The indices $s$ and $\lambda$ run through the 4 Stokes
parameters and the wavelength, respectively. Note that the merit
function inside the STiC code has additional terms that apply the
regularization scheme and the normalization of $\chi^2$.


\section{Analysis}\label{sec:analysis}

\subsection{Disk center, $\mu = 1$}
\begin{figure}[!t]
\includegraphics[angle=0,scale=.47]{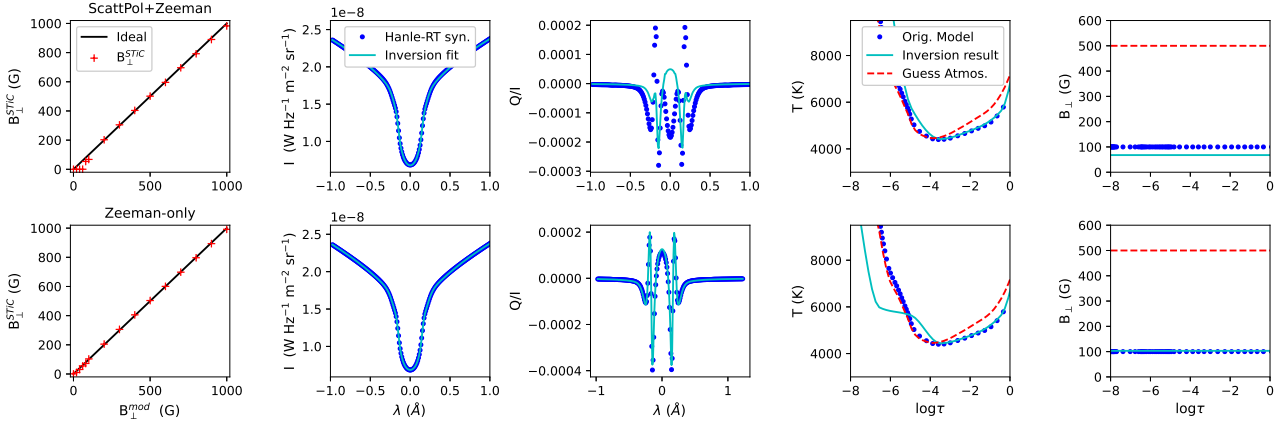}
\caption{Inversion results for the LOS with $\mu=1$. The top row shows results for
  the cases in which atomic polarization was accounted for, while the
  bottom row concerns itself with the Zeeman-only case. The first
  column shows the retrieved magnetic field vs. the model value, where
  the solid line represents the one-to-one expected correspondence. The
  second and third columns show the fits (cyan lines) to the Hanle-RT
  Stokes I and Q/I (blue dots) for the case of a magnetic field
  of $\sim 100$ G. The last two
  columns present the retrieved temperature and magnetic field (cyan
  line) as a function of optical depth, and how they compare to the model used for the Hanle-RT
  synthesis (blue dots), as well as the initial guess for the
  inversion (red line).
  \label{fig:mu10}}
\end{figure}

Figure \ref{fig:mu10} summarizes the results for the observing
geometry with $\mu=1.0$. The top row shows inversion results for the
profiles computed with scattering polarization and the joint action of
the Hanle and Zeeman effects, whilst the bottom row presents the
results for the Zeeman-only case.
The left-most column compares the values of the horizontal field
retrieved by the STiC inversion, B$_{\perp}^{\rm STiC}$, to the
field in the model, B$_{\perp}^{\rm mod}$ (with
which the synthetic Hanle-RT profiles were generated). 
The retrieved magnetic field values in the
Zeeman-only case (bottom) are always extremely close to the truth
(black solid line).  The initial
atmospheric guess as well as the choice of weights for the Stokes
parameters impact the solution to some degree, resulting in very small
deviations from the truth.
For the case of the full calculation (top) the agreement is
remarkable and not qualitatively different to the Zeeman-only case
when the field strength is above $\sim 100$ G. Significant disagreements
between the retrieved field values and the expected ones appear below
this threshold. In this regime, the signatures of the Hanle effect
alter the shape of the linear polarization profiles significantly,
resulting in a misinterpretation from the STiC code.

\noindent The second and third columns show the best fits
found by the inversion (cyan line) to the Hanle-RT Stokes I and 
Q/I (blue dots),
synthesized for a magnetic field of 100 G. In this particular geometry,
U/I and V/I are exactly zero.
It is obvious that the STiC inversions are not able to fit the signatures
characteristic of the Hanle effect (Q/I panel in top row), while it does a good job at fitting the
Zeeman-only profile (bottom row). Despite the misfit of the former,
STiC is still able to infer an approximate value of the magnetic
field by virtue of attempting to fit the width and amplitude of the
Q/I profile. It is worth noting that, in general, the quality of
the fits becomes increasingly better as the field strength
increases and the Hanle contribution becomes less and less important. Below 100 G,
the inversion fits are poor and the inferred magnetic field
values are rather inaccurate.

\noindent Columns 4 and 5 compare the
retrieved temperature and horizontal magnetic field
stratifications as a function of optical depth (solid cyan lines) to the model
atmosphere used for the synthesis (blue dots). The disagreements
between the original and the inverted temperature profiles are partly 
due to a degeneracy between this physical quantity and the
microturbulent velocity. This degeneracy can be mitigated by carrying
out simultaneous inversions
of multiple spectral lines with different sensitivities to temperature
\cite[see, for instance,][]{stic, silva2018}. That said, the discrepancies are larger at 
optical depths
where Ca {\sc ii} 8542 \AA\ lacks sensitivity to the temperature,
especially above ${\rm log}\tau = -5.5$ \cite[][]{quinteronoda}.
In the regions of the atmosphere where the Ca {\sc ii} 8542 \AA\ line
should have sensitivity to the temperature, some of the discrepancy
between the original model and the inversion results comes from the fact that
the Hanle-RT syntheses were carried out in the FALC model atmosphere with
electron density and hydrogen populations calculated in non-LTE. For
the inversion, on the other hand, the STiC code was set up to solve the statistical equilibrium for the Ca
II atom only, while the H populations and electron densities were computed
internally under the assumption of LTE.
Additional sources of discrepancy, more difficult to quantify, stem
from the differences between the Hanle-RT and STiC codes in the formulation of the equation of state, the
algorithm that computes the formal solution of the equation of
transfer, and other factors that might affect the convergence of the statistical
equilibrium.
In these last two columns of Fig. \ref{fig:mu10}, the red line shows the guess model atmosphere used to initialize the inversion.

\subsection{Other geometries}

\begin{figure}[!t]
\includegraphics[angle=0,scale=.58]{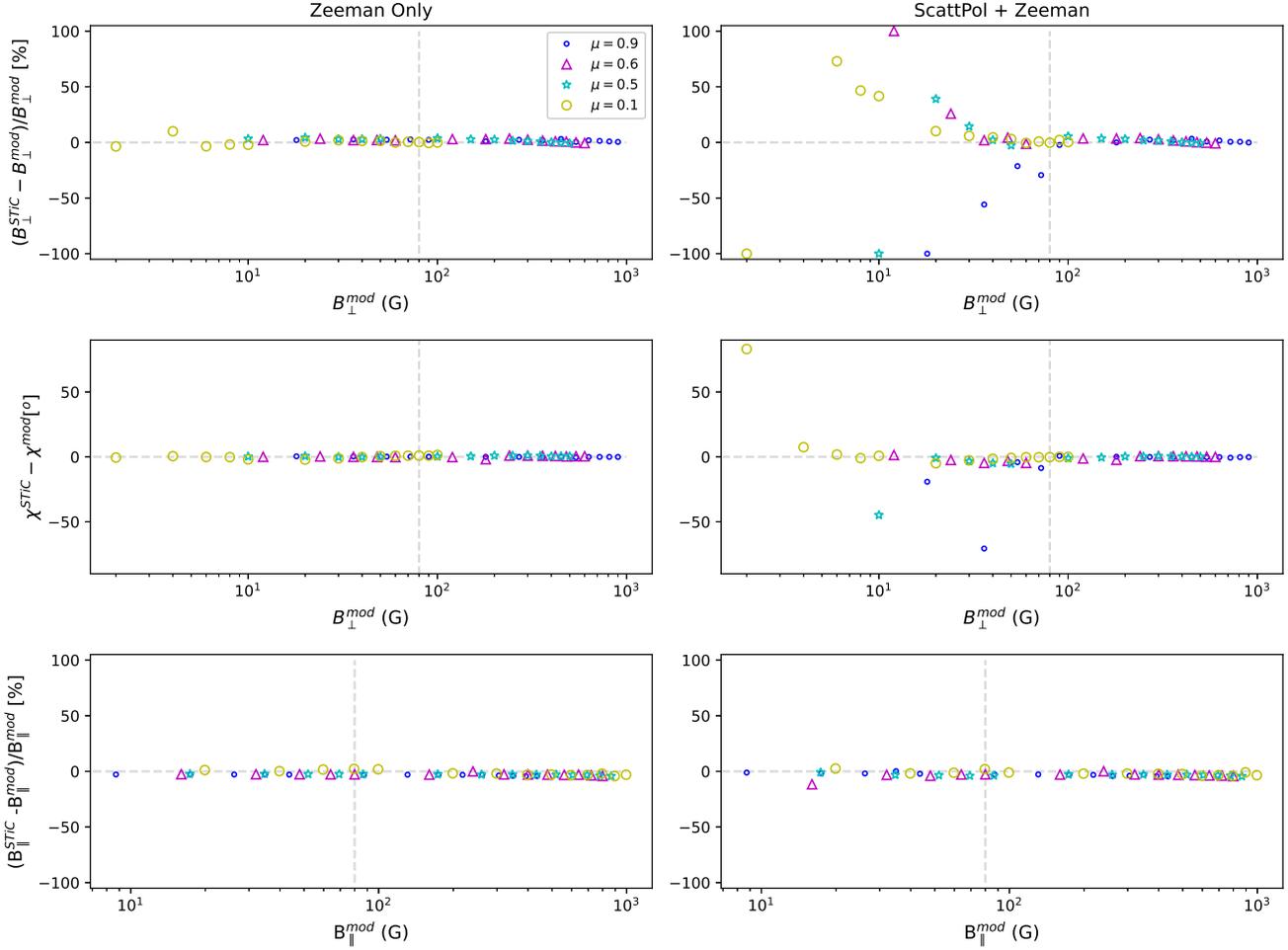}
\caption{Inversion results for observing geometries away from the disk
center. The panels show the relative (percent) differences between
the magnetic field values retrieved by the STiC inversions and those
in the model atmospheres used for the Hanle-RT syntheses. Left and right
correspond, respectively, to the Zeeman-only and the full calculation
with atomic polarization cases. The top two rows show the results for
the strength and the azimuth of the transverse
component of the magnetic field, while the bottom panels present the
results for the LOS magnetic field.
  \label{fig:mu_other}}
\end{figure}

As we move to observing geometries away from the disk center, the magnetic field
presents both a longitudinal as well as a transverse
component in the LOS reference frame. This leads to the appearance of
Stokes V signals exclusively due to the longitudinal Zeeman effect. Furthermore, the
fraction of Zeeman-induced linear polarization becomes smaller and
the effects of scattering polarization become more important as the
heliocentric angle of the ``observation'' increases (i.e. $\mu$ decreases).

Figure \ref{fig:mu_other} shows a comparison between
the components of the vector magnetic field retrieved by the STiC inversions and
those used for the Hanle-RT syntheses, for the LOS away from
the disk center. The left column corresponds to the
inversion of Zeeman-only profiles whilst the right column depicts the
results for the full forward calculation with atomic polarization.

The top row quantifies the relative (percent) difference between the
inferred (${\rm B}_\perp^{\rm STiC}$) and model (${\rm B}_\perp^{\rm
  mod}$ ) values of the transverse magnetic field, as a function of the
model value. The
x-axis is on a logarithmic scale to allow the reader to clearly see
the results at low field strengths (${\rm B}_\perp^{\rm mod} <100$~G),
and each line of sight is represented by a colored symbol.
For transverse magnetic field strengths above 80 G, the inversion
results are always within 5\% of the true
value, regardless of whether STiC is interpreting the spectra from the
full scattering polarization calculation or the Zeeman-only signals. Below 80 G (marked by the vertical
gray line), the inferences derived from the spectra with scattering
polarization (right) start to deviate significantly from the true
values, whilst in the case of the Zeeman-only signals (left), the inversion
inferences are still accurate.
 The middle row shows the difference between the inferred and the
 model magnetic field azimuths (in degrees) as a function of ${\rm B}_\perp^{\rm mod}$.
Just like in the case of the horizontal field strength, the inferences from the
scattering polarization signatures become less reliable below the 80 G mark.

The bottom row presents the inversion results
for the longitudinal component of the magnetic field, ${\rm
  B}_\parallel^{\rm STiC}$, and how they
compare to the model values (${\rm
  B}_\parallel^{\rm mod}$). Because the atomic level polarization only
carries atomic alignment, it does not affect the circular polarization
signals, which are exclusively due to the Zeeman effect. Therefore,
the Stokes V signals are correctly
interpreted by STiC, and the ${\rm B}_\parallel^{\rm STiC}$ inferences
are always rather accurate.


\subsection{The case of $\mu=0.1$}

Special consideration should be given to the case of the LOS with $\mu = 0.1$. This
corresponds to the observing geometry closest to the limb, and
therefore has the smallest transverse 
magnetic fields in the observer's reference frame (${\rm B}_{\perp}^{\rm mod}$ ranging from 0 to $100$~G). Also, the linear polarization profiles in
this scenario will be the most affected by scattering polarization and
the Hanle effect.

\begin{figure}[!t]
\includegraphics[angle=0,scale=.47]{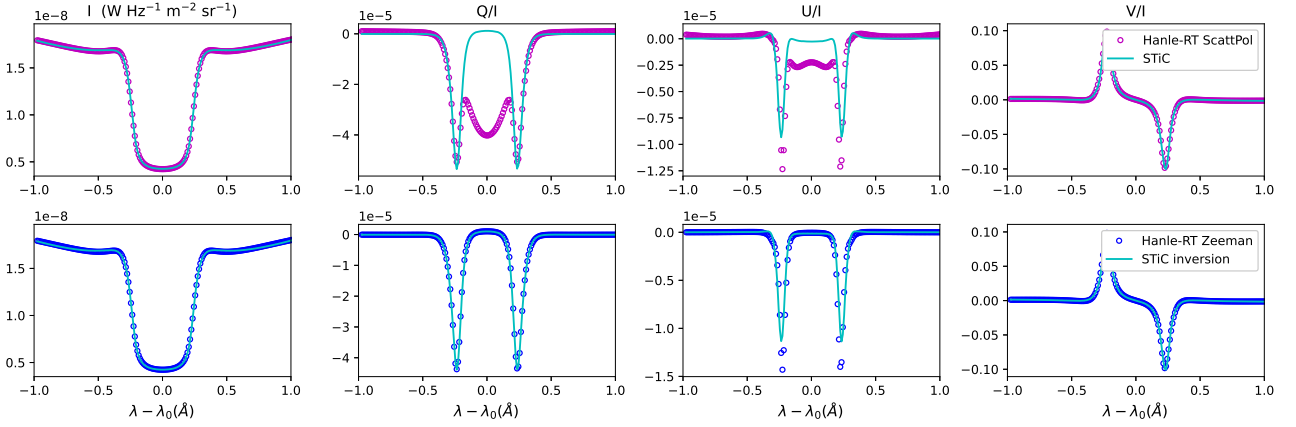}
\caption{Inversion fits for the case of ${\rm B}^{\rm mod}= 200$~G and observing
  geometry with $\mu = 0.1$. The open circles represent the Hanle-RT
  synthetic profiles, while the cyan line shows the best fit found by
the inversion code. The top row shows the case of the full calculation
while the bottom row corresponds to the Zeeman-only spectra.}
  \label{fig:limb}
\end{figure}

Figure \ref{fig:limb} shows the best inversion fits for the case
${\rm B}^{\rm mod}=200$~G in the observing geometry $\mu=0.1$. In the LOS reference frame, this corresponds to a transverse
magnetic field of ${\rm B}_{\perp}^{\rm mod}=20$~G and a LOS component
of ${\rm B}_{\parallel}^{\rm mod}=199$~G.
 The top row corresponds
to the full calculation, including scattering polarization and the
Hanle effect, while the bottom row shows the results for the
Zeeman-only calculations. The open circles represent the synthetic
profiles calculated with Hanle-RT (from left to right, $I$, $Q/I$, $U/I$,
$V/I$) while the cyan solid line shows the best fit found by the STiC inversion
code. 
\noindent This figure epitomizes how the inversion code interprets the signatures
introduced by atomic polarization and the Hanle effect in the linear polarization profiles.
In particular, the second panel of the top row shows the Stokes $Q/I$
profile affected these subtle quantum mechanical effects
(magenta circles), which is to be compared to its Zeeman-only
counterpart in the bottom row (blue circles). In both cases, the
best fits found by the inversion code (cyan lines) look rather similar, and yield
comparable values of the transverse magnetic field: the inversions of
the Zeeman-only case yields $20.2$~G, whilst the full calculation
results in $22.0$~G. The model value was
${\rm B}_{\perp}^{\rm mod}=20$~G in this case. 
Even though the Hanle effect strongly depolarizes the $\pi$ component of
the Stokes $Q/I$ signal, there might be enough information in the inner wings
for STiC to extract information about
the magnetic field through the Zeeman effect. The resulting
${\rm B}_{\perp}^{\rm STiC}$ is only
over-estimated by 10\% in this case.

\subsection{The zero-field case}
\begin{figure}[!t]
\includegraphics[angle=0,scale=.55]{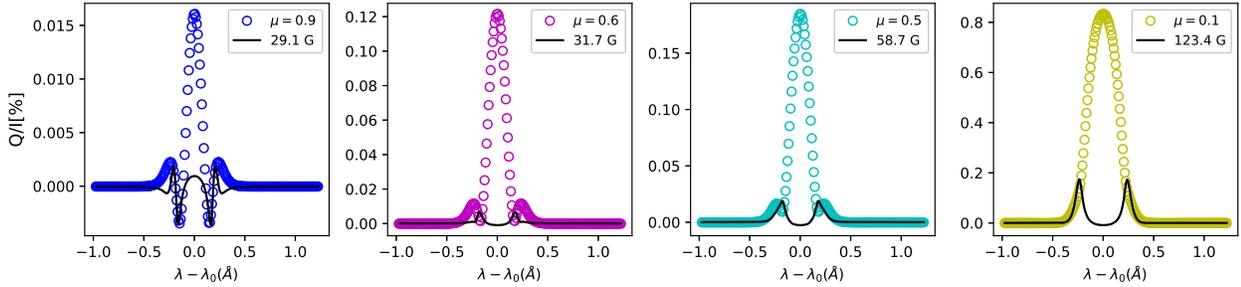}
\caption{Inversion fit (black line) of $Q/I$ for the zero-field
  case when scattering polarization is accounted for (open circles). STiC attempts to fit
  the non-magnetic scattering polarization signals as if they were of
  Zeeman origin, leading to non-zero inferences
  of the transverse component of the magnetic field (the value
  retrieved in each case is listed in the legend).}
  \label{fig:zero_field}
\end{figure}

The case of the full calculation for ${\rm B}=0$~G misleads the
inversion code, since the purely
scattering polarization Stokes $Q$ signals are interpreted by STiC as Zeeman-induced
transverse magnetic fields of tens to a hundred gauss.
Figure \ref{fig:zero_field} shows the STiC fits (black solid lines) to the
Hanle-RT $Q/I$ profiles (open circles) for the case of the full calculation
with ${\rm B}=0$~G. From
left to right, the observing geometry goes from $\mu=0.9$ to
$\mu=0.1$. As the figure suggests, STiC is not able to fit the 
$Q/I$ profiles, yet it attempts to mimic the scattering polarization signals in order to minimize the merit
function that drives the inversion algorithm. This leads to non-zero values of
the magnetic field, which in the case of the most extreme observing geometry
yields ${\rm B}_{\perp}^{\rm STiC} \sim 120$~G.

\section{Conclusions}\label{sec:conclusions}

Most spectral line inversion codes that are able to interpret 
chromospheric spectra formed under non-LTE conditions do not account for all the
physical processes that can generate polarization in these lines.
Although the Zeeman effect is taken into account in these codes 
\cite[][]{stic, nicole, snapi, desire}, the physics for
the generation and transfer of polarization due to scattering and its modification via the
Hanle effect are not.
This work assesses the errors made by one of these inversion codes (STiC) when
interpreting the polarization signatures of Ca {\sc ii} 8542 \AA.

To estimate the errors incurred by STiC when interpreting the Ca {\sc
  ii} 8542 \AA\ spectra we invert sets of Stokes profiles synthesized
with the Hanle-RT code in the FALC model atmosphere with ad-hoc magnetic
fields of different strengths. 
Hanle-RT computes the spectral line polarization generated by
scattering processes as well as its modification due to
the combined actions of the Hanle and Zeeman effects. 
The spectra were synthesized simulating 5 observing geometries, from disk
center ($\mu=1$) to close to the solar limb ($\mu=0.1$).

We find that, for most field strengths and observing geometries, STiC
does a good job at retrieving both the longitudinal and the transverse
components of the magnetic field.
Even when the amplitude of the scattering polarization and Hanle
signatures are comparable to that of the Zeeman signature in the
linear polarization profiles, STiC is able to retrieve the transverse component of the
magnetic field with remarkable accuracy, over-estimating its value by
only a few percent. 
When the transverse magnetic field component is weaker than ~80 G, the
scattering polarization and Hanle signatures start dominating over the
Zeeman-induced signals. Naturally, STiC is not able to reproduce
the spectral signatures induced by scattering polarization and the
Hanle effect, leading to errors in the retrieved transverse component of the
magnetic field and its azimuth in the plane of the sky. 

In the case where the model magnetic field is ${\rm B}^{\rm mod}=0$~G, and
polarization signatures are exclusively due to scattering processes,
STiC still attempts to fit the linear polarization by interpreting it as a
Zeeman signal. This leads to spurious transverse magnetic field
values, which in the case of $\mu=0.1$ is just over $120$~G. However, by
visually inspecting the original spectra as well as the
inversion fits, it is possible to determine that the agreement is poor and that
the inversion results are questionable.

This is a relatively simple experiment that only addresses spectral profiles
synthesized in a 1-D static semi-empirical model atmosphere with constant magnetic
fields in the absence of noise in the Stokes parameters. {\bf 
The 1-dimensional nature of this study omits, by design,
the effects of spatial inhomogeneities and horizontal
radiative transfer on the atomic level polarization. This likely leads
to an under- or over-estimate of the typical linear polarization
signals expected in very quiet areas of the (real) Sun, which will become less
important in more strongly magnetized regions.}
Moreover, as mentioned in the introduction,
strong velocity gradients can lead to much larger scattering
polarization signatures in observing geometries close to the limb. The
interpretation of such signals with a non-LTE inversion code could
lead to larger errors in the retrieved magnetic field values.
Nevertheless, this work provides a first attempt at setting a threshold for
the applicability of non-LTE spectral line inversion codes, and shows
that, when the contribution of scattering polarization to the linear
polarization signals is smaller than that of its Zeeman-induced counterpart, the
inversion results are as accurate as when the polarization signals are due to
the Zeeman effect only.


This material is based upon work supported by the National Center for
Atmospheric Research, which is a major facility sponsored by the
National Science Foundation under Cooperative Agreement
No. 1852977. This project has received funding from the European
Research Council (ERC) under the European Union's Horizon 2020
research and innovation programme (SUNMAG, grant agreement
759548). TdPA acknowledges the funding received from the European
Research Council (ERC) under the European Union’s Horizon 2020 research and innovation programme (ERC Advanced Grant agreement No 742265).




\end{document}